# Hydro-Gravitational-Dynamics of Planets and Dark Energy

Carl H. Gibson [1,2]

[1] University of California San Diego, La Jolla, CA 92093-0411, USA
[2] cgibson@ucsd.edu, http://sdcc3.ucsd.edu/~ir118

and

Rudolph E. Schild[3,4]

[3] Center for Astrophysics, 60 Garden Street, Cambridge, MA 02138, USA
[4] rschild@cfa.harvard.edu

**Abstract:** Self-gravitational fluid mechanical methods termed hydro-gravitational-dynamics (HGD) predict plasma fragmentation 0.03 Myr after the turbulent big bang to form protosuperclustervoids, turbulent protosuperclusters, and protogalaxies at the 0.3 Myr transition from plasma to gas. Linear protogalaxyclusters fragment at 0.003 Mpc viscous-inertial scales along turbulent vortex lines or in spirals, as observed. The plasma protogalaxies fragment on transition into white-hot planet-mass gas clouds (PFPs) in million-solar-mass clumps (PGCs) that become globular-star-clusters (GCs) from tidal forces or dark matter (PGCs) by freezing and diffusion into 0.3 Mpc halos with 97% of the galaxy mass. The weakly collisional non-baryonic dark matter diffuses to > Mpc scales and fragments to form galaxy cluster halos. Stars and larger planets form by binary mergers of the trillion PFPs per PGC on 0.03 Mpc galaxy accretion disks. Star deaths depend on rates of planet accretion and internal star mixing. Moderate accretion rates produce white dwarfs that evaporate surrounding gas planets by spin-radiation to form planetary nebulae before Supernova Ia events, dimming some events to give systematic distance errors misinterpreted as the dark energy hypothesis and overestimates of the universe age. Failures of standard ΛCDM cosmological models reflect not only obsolete Jeans 1902 fluid mechanical assumptions, but also failures of standard turbulence models that claim the cascade of turbulent kinetic energy is from large scales to small. Because turbulence is always driven at all scales by inertial-vortex forces $\vec{v} \times \vec{\omega}$ the turbulence cascade is always from small scales to large.

## 1. Introduction

Dimness of supernovae Ia (SNe Ia) events for redshift values 0.01 < z < 2 have been interpreted as an accelerating expansion rate for the universe [1-3]. The acceleration is attributed to mysterious antigravity effects of “dark energy” and a cosmological constant Λ. Dimming is observed at all frequencies by about 30%, with large scatter attributed to uncertainty in the SNe Ia models. Hubble Space Telescope Advanced Camera for Surveys (HST/ACS) images have such high signal to noise ratios that both the scatter and the dimming are statistically significant over the full range of z values. Bright SNe Ia observed for z ≥ 0.46 exclude “uniform grey dust” systematic errors, supporting flat-universe deceleration until the recent “cosmic jerk” to acceleration for z ≤ 0.46. The “dark energy” interpretation is a consequence of the commonly accepted Λ-cold-dark-matter (ΛCDM) cosmological theory. However, ΛCDM theory is fluid mechanically untenable. The theory assumes irrotational, collisionless and frictionless flows and neglects viscosity, turbulence[1], diffusion, fossil turbulence and fossil turbulence waves. Hydro-gravitational-dynamics (HGD) is the application of modern fluid mechanics to cosmology [11-16]. As predicted by HGD, primordial planets in protoglobularstarcluster clumps formed at $z = 1100$, $t = 10^{13}$ s comprise the dark matter of galaxies and the source of all stars. Because dying stars may be dimmed by gas planets evaporated near the star, planets provide an alternative to dark energy. New physical laws are not required by HGD but ΛCDM must be discarded. The choice is between planets and dark energy.

[1] Turbulence [4,5] is defined as an eddy-like state of fluid motion where the inertial-vortex forces of the eddies are larger than any other forces that tend to damp the eddies out. Turbulence by this definition always cascades from small scales to large, starting at the viscous-inertial-vortex Kolmogorov scale at a universal critical Reynolds number. The mechanism of this «inverse» cascade is merging of adjacent vortices with the same spin due to induced inertial vortex forces that force eddy mergers at all scales of the turbulence. Such eddy mergers account for the growth of turbulent boundary layers, jets and wakes. A myth of turbulence theory is that turbulence cascades from large scales to small. It never does, and could not be universally similar if it did. Irrotational flows cannot be turbulent by definition. In self-gravitational fluids such as stars and planets, turbulence is fossilized in the radial direction by buoyancy forces. Radial transport is dominated by fossil turbulence waves and secondary (zombie) turbulence and zombie turbulence waves in a beamed, radial, hydrodynamic-maser action.

The definition of turbulence and its implied small-to-large cascade direction for turbulent energy transfer [5] matches that used [4] to extend the well-known Kolmogorov universal similarity laws of turbulence to stably stratified turbulent mixing of scalar fields like temperature and electron density with variable Prandtl number $\mathrm{Pr} = \alpha / \nu$, where $\alpha$ is thermal diffusivity and $\nu$ is kinematic viscosity [6-8]. Substantial evidence supports these similarity laws, the universal constants of second order structure functions and energy spectra, as well as intermittency constants that arise with " Kolmogorov third law" refinements despite claims to the contrary [9]. The direction of the turbulent energy cascade and the definition of turbulence depend on the Navier-Stokes equations written so the nonlinear $\vec{v} \times \vec{\omega}$ term causing turbulence is isolated ($\vec{\omega} = \nabla \times \vec{v}$)

$$\partial \vec{v} / \partial t = \nabla B + \vec{v} \times \vec{\omega} + \vec{F}_{Viscous} + \vec{F}_{Buoyancy} + \vec{F}_{Coriolis} + \vec{F}_{Other} \qquad (1)$$

where $\vec{v}$ is the fluid velocity, B is the Bernoulli group of mechanical energy terms $v^2 + p/\rho + lw$, and the various fluid forces listed tend to damp out turbulent motions driven by $\vec{v} \times \vec{\omega}$. The kinetic energy per unit mass $v^2$, the specific enthalpy $p/\rho$ and frictional lost work lw in B are generally constant along streamlines in natural hydrodynamic flows so their gradient can be neglected. The ratio of $\vec{v} \times \vec{\omega}$ to the viscous, buoyancy and Coriolis forces defines the Reynolds, Froude and Rossby numbers, respectively. For a flow to be turbulent, universal critical values of Re, Fr and Ro must be exceeded. The direction of the turbulence kinetic energy cascade is not essential to most laboratory studies, but it is crucial to natural flows where fossil turbulence[2] and fossil turbulence waves are important to preserve information about previous turbulence and turbulent fluxes and as components of the dynamical transport mechanism. Most temperature and salinity microstructure in the ocean is fossil turbulence because no mechanism besides molecular diffusion exists to erase these signatures of previous turbulence, even though the kinetic energy of the turbulence at the scale of the fossils has long since radiated away as internal wave motions, termed fossil turbulence waves.

Fossil turbulence patches are identified by means of hydrodynamic phase diagrams $\mathrm{Fr}_{Patch}$ versus $\mathrm{Re}_{Patch}$ [9]. It is shown that

$$\mathrm{Fr}_{\mathrm{Patch}} = \mathrm{Fr}/\mathrm{Fr}_0 = \left(\varepsilon / \varepsilon_0\right)^{1/3} = \left(\varepsilon / 3 L_{T0}{}^2 N^3\right)^{1/3} \qquad (2)$$

and

$$\mathrm{Re}_{Patch} = \varepsilon / \varepsilon_F = \varepsilon / 35 \nu N^2, \qquad (3)$$

where dissipation rates $\varepsilon$ are normalized by Fr and Re values estimated by fossil turbulence theory assuming the cascade of turbulence is from small scales to large. From (2) we see that the Thorpe overturning scale $L_{T0}$ at the beginning of fossilization in a fossilized microstructure patch may be much larger than the overturning scale at the time of sampling [10]. Without a turbulence definition based on $\vec{v} \times \vec{\omega}$ and a turbulence cascade from small scales to large the signatures of fossil turbulence are not unique. Thousands of hydrodynamic phase diagrams are available from oceanic microstructure studies that demonstrate the dynamics of fossil turbulence formation, contradicting the common oceanographic assumption that fossil turbulence does not exist and the common turbulence assumption that turbulence cascades from large scales to small [4,6,9,10].

---

[2] Fossil turbulence [4,6,9,10] is a perturbation in any hydrophysical field caused by turbulence that persists after the fluid is no longer turbulent at the scale of the perturbation.

Turbulent kinetic energy dissipates to heat at viscous Kolmogorov scales $L_K = \left(\nu^3/\varepsilon\right)^{1/4}$, so it is reasonable to assume that it cascades down to this scale from some larger scale where it originates. This is the standard turbulence model, expressed by the poem of L. F. Richardson 1922 in all turbulence textbooks. However, the flows from which turbulence extracts energy are irrotational at large scales with $\vec{v}\times\vec{\omega}$ nearly zero, so they are non-turbulent by definition. Turbulent kinetic energy is not only dissipated at the Kolmogorov scale, but is also produced at $L_K$. Adjacent eddies with the same spin induce $\vec{v}\times\vec{\omega}$ forces on each other in directions that cause the eddies to merge, which is the physical mechanism of the small-to-large turbulent energy cascade.

An example is merging of eddies formed in a boundary layer, which grow and induce $\vec{v}\times\vec{\omega}$ forces away from the boundary layer causing boundary layer separation. Adjacent eddies with opposite spin induce translational and divergence forces that cause ingestion of irrotational external flows into the interstices of the turbulence down to viscous scales where they acquire spin and can be classified as turbulence. In natural flows such as the ocean, atmosphere and galaxy, turbulent flows are limited by buoyancy, Coriolis and other forces at large scales. Most of the turbulent kinetic energy produced at Kolmogorov scales is not dissipated as heat but is radiated or stored as waves. Information about previous turbulence events is stored by a variety of hydrophysical fields such as temperature and electron density by fossil turbulence remnants. Because turbulence events in natural flows are quite brief compared to the persistence time of fossil turbulence, most of the microstructure of the ocean, atmosphere and galaxy are turbulent fossils.

Turbulence is absolutely unstable. A vortex sheet thickens to a Kolmogorov scale $L_K$ starting from zero in a Kolmogorov time $T_K = \left(\nu/\varepsilon\right)^{1/2}$ when the first eddies appear. The cascade of turbulent kinetic energy is to larger scales because the overturn time $T_{Overturn} = L^{2/3}\varepsilon^{-1/3}$ increases with the overturn scale L (from Kolmogorov's second similarity hypothesis). Similarly, self gravitation is absolutely unstable [12]. Starting from a mass perturbation $M'(t_0)$ at the origin of a motionless large body of uniform density fluid with no forces other than gravity, the mass will grow or decrease with time depending on the sign of the perturbation and the gravitational free fall time $\tau_g = \left(\rho G\right)^{-1/2}$, where G is Newton's gravitational constant and $\rho$ is the density. The perturbation

$$M'(t) = \left|M'(t_0)\right| \exp\left[\pm 2\pi\rho G t^2\right] = \left|M'(t_0)\right| \exp\left[\pm 2\pi\left(t/\tau_g\right)^2\right] \qquad (4)$$

suddenly grows exponentially to infinity with time if the perturbation $M'(t_0)$ is positive. If the perturbation $M'(t_0)$ is negative the density suddenly decreases exponentially to zero forming a void at the free fall time. Before either of these singularities are reached, viscous forces or turbulence forces or molecular diffusivity effects will appear to cushion the gravitational collapse or fragmentation. The effect will occur at the largest of three Schwarz scales [11,12]; that is, at

$$L = L_{SX} = \max\left[L_{SV}, L_{ST}, L_{SD}\right] \qquad (5)$$

where $L_{SV} = \left(\nu\gamma/\rho G\right)^{1/2}$, $L_{ST} = \left(\varepsilon/\left[\rho G\right]^{3/2}\right)^{1/2}$, and $L_{SV} = \left(D^2/\rho G\right)^{1/4}$.

A simple demonstration of the direction of the turbulence cascade is to fill a bottle completely with water with a powder (say paprika) to show the small scale motions. It is impossible to make the interior fluid turbulent without spinning the bottle. Large scale motions have no effect, but spin causes turbulence to form from shears at the solid-liquid interface and the turbulence cascade from small scales to large fills the bottle with turbulent motions. The spectral cascade is illustrated in Fig. 0, where nearly motionless fluid becomes turbulent by spinning the bottle at constant angular velocity. Initially the turbulent kinetic energy spectrum $\phi(k)$ has a small range of wavenumbers $k$ and small mean-square velocity integral $\int \phi(k) dk = \langle v^2 \rangle$, all in a narrow range $k_2 > k_{BL2} = (5L_{K2})^{-1}$. The turbulence of the boundary layer cascades to the bottle scale $L_{bottle} > 5L_K$ at stage 4 and then is damped by viscosity in stages 5 and 6 as the fluid reaches solid body rotation.

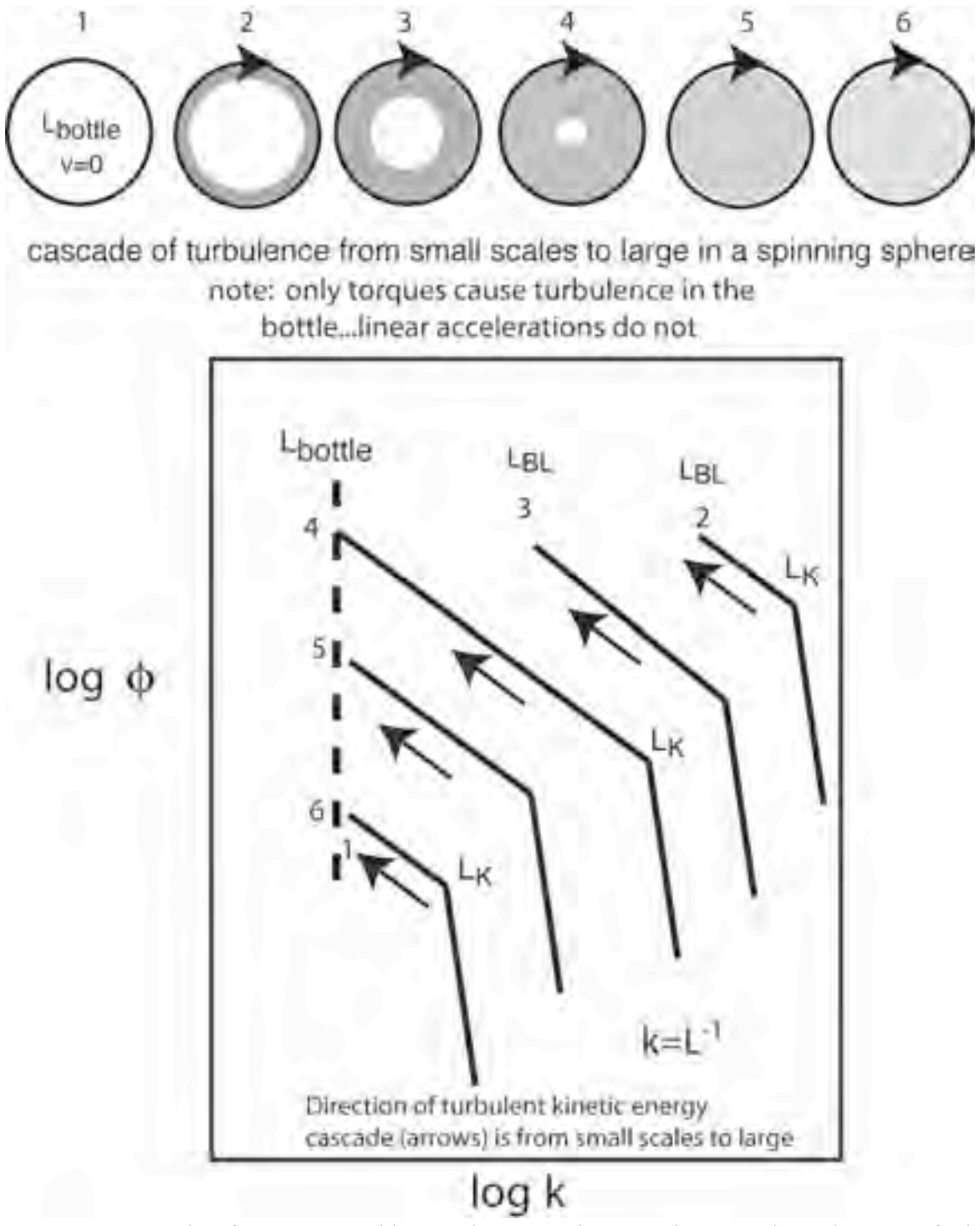

Fig. 0. Demonstration of the turbulence cascade from small scales to large by spinning a full bottle of fluid from rest with constant angular velocity. Turbulence forms at the boundaries of the bottle starting with $L_{BL2} \approx 5L_K$ and grows till the larges scale of the turbulence matches the bottle size. The direction of the energy cascade is shown by arrows.

The most extreme example of the turbulence cascade and turbulence fossilization is that of big bang turbulence [13,14], where turbulence begins at the Planck scale $10^{-35}$ m and cascades to $10^{-27}$ m where the turbulent temperature fluctuations are fossilized by inflation stretching them beyond the scale of causal connection $L_H = ct$.

In the following Section 2 we discuss the theories of gravitational structure formation, followed by a review of the observational data in Section 3, a Discussion of Results in Section 4 and finally some Conclusions in Section 5.

## 2. Theory

The Jeans 1902 linear theory of acoustic gravitational instability [15] predicts only one criterion for structure formation. Fluctuations of density are assumed unstable at length scales larger than the Jeans length $L_J = V_S / (\rho G)^{1/2}$ but stable for smaller scales, where $V_S$ is the speed of sound, $\rho$ is density, $G$ is Newton's gravitational constant, and $(\rho G)^{-1/2}$ is the gravitational free fall time. Because the speed of sound in the plasma epoch after the big bang is nearly the speed of light, the Jeans scale for the plasma is always larger than the scale of causal connection *ct*, where *c* is the speed of light and *t* is the time since the big bang, so no gravitational structures can form in the plasma. Jeans' 1902 fluid mechanical model [15] is the basis of ΛCDM, where an unknown form of collisionless non-baryonic dark matter (NBDM) is assumed to condense because it is somehow created cold so its sound speed and Jeans length can be assumed smaller than *ct*. The NBDM clumps cluster hierarchically and magically stick together to form potential wells in ΛCDMHC models [1, 2] with dark energy. CDM clumps have been sought but not observed, as predicted by HGD.

It is not true that the primordial plasma is collisionless and it is not true that gravitational instability is linear. Gravitational instability is intrinsically nonlinear and absolute in the absence of viscosity just like the inertial-vortex-force $\vec{v} \times \vec{\omega}$ turbulence instability is nonlinear and absolute in the absence of viscosity. All fluid density fluctuations are unstable to gravity, and in the plasma epoch $10^{11} \le t \le 10^{13}$ s structures will form by gravitational forces at all scales less than *ct* unless prevented by diffusion, viscous forces or turbulence forces, as shown in Figure 1 [11].

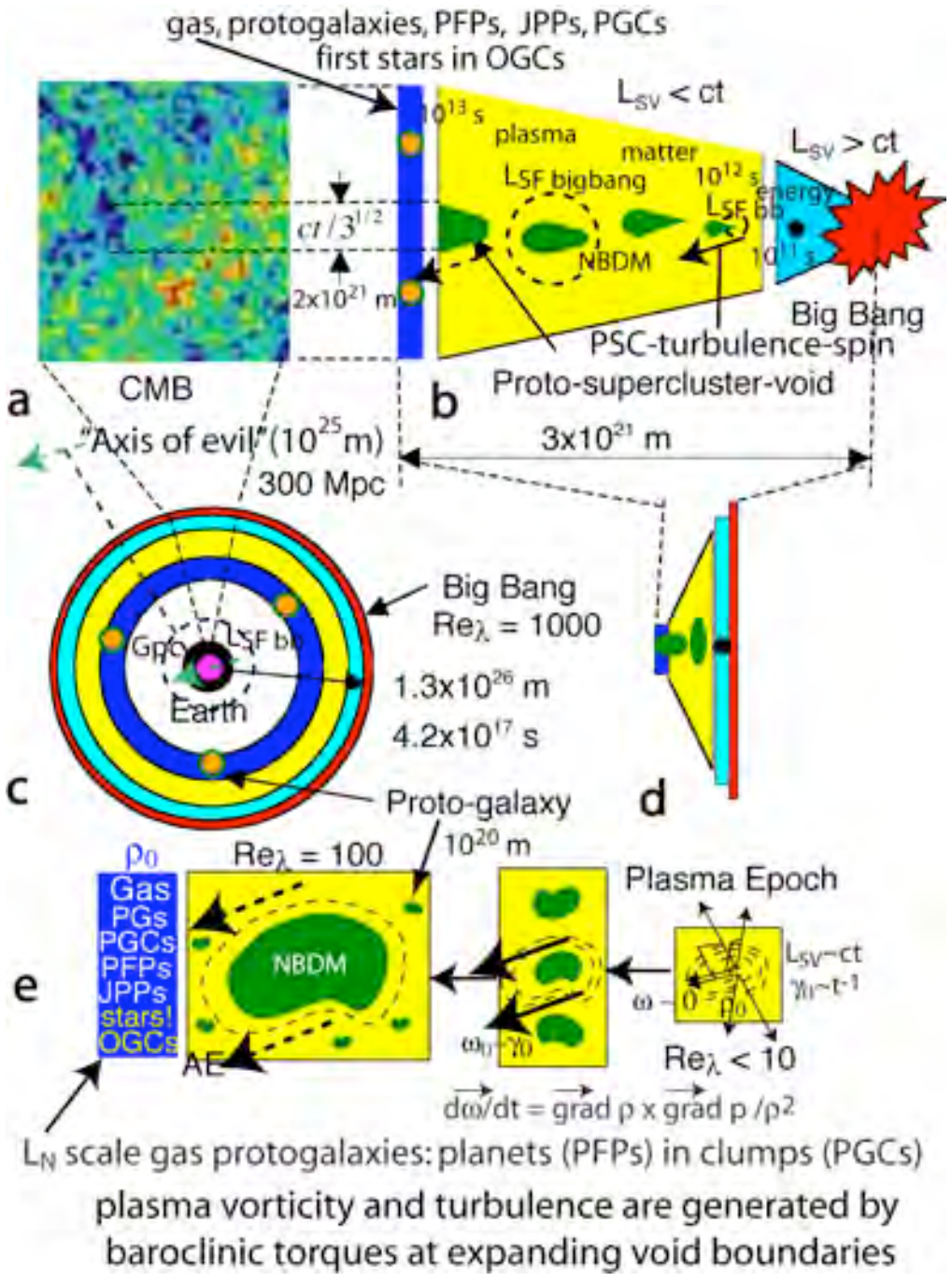

Fig. 1. Formation of gravitational structures according to hydro-gravitational-dynamics (HGD). The entire baryonic plasma universe fragments at the gas Schwarz viscous scale to form planets in Jeans mass clumps in Nomura geometry [9] proto-galaxies at transition. Plasma fossil big bang turbulence density gradients produce "axis of evil" quasar, galaxy and galaxy cluster spin alignments to Gpc scales [11, 22] by baroclinic torques at the expanding void boundaries.

The kinematic viscosity of the plasma is determined by photon collisions with free electrons that drag their protons and alpha particles with them. The photon viscosity of the plasma epoch $\nu_p \approx 4 \times 10^{26}$ m$^2$ s$^{-1}$ [5] is so large that the potential wells of CDM halos have Reynold numbers less than one and could not fill with plasma in the available time, and any accoustic oscillations from the filling would be prevented by photon-viscosity. Thus the dominant accoustic peak of the CMB temperature anisotropy spectrum of Fig. 1a is explained by HGD theory from the sonic-speed expansion of protosupercluster voids triggered by plasma density minima as shown in Fig. 1b, but not by CDM theory if viscosity is included in the numerical simulations. The mean free path for photon-electron collisions is less than the horizon scale by two orders of magnitude at the time of first structure from the Thomson scattering cross section and the known electron density, as required by the continuum hypothesis of fluid mechanics.

When the Schwarz viscous-gravitational scale $L_{SV}$ becomes less than the horizon scale $L_H = ct$ then gravitational structure formation begins. The first structures are fragmentations because the rapid expansion of the universe inhibits condensation at density maxima but unhances void formation at density minima. It is a myth of astrophysics that pressure-support or thermal-support will prevent structure formation at scales smaller than the Jeans scale $L_J = V_S / (\rho G)^{1/2}$. Pressure support occurs in hydrostatics, not hydrodynamics. At the plasma-gas transition the kinematic viscosity decreases by a factor of $\approx 10^{13}$, permitting fragmentation at gas-planetary rather than plasma-galactic scales $L_{SV} = (\gamma\nu / \rho G)^{1/2}$. Viscous forces are no longer able to prevent void growth at density minima so the protogalaxies fragment to form primordial fog particles. From the Bernoulli equation the pressure decreases during the fragmentation process as the specific enthalpy $p / \rho$ decreases to compensate for the increasing kinetic energy per unit mass $v^2/2$ along streamlines toward density maxima or away from density minima. Consider what happens when a cannonball or a vacuum beachball suddenly appears in a uniform motionless gas [5]. In the first case gravity accelerates everything toward the center. In the second case gravity accelerates everything away. Nothing much happens until the gravitational free fall time, at which the density exponentiates to infinity or zero.

In either case the pressure gradient is in the direction of motion, opposite to that required for the fictitious pressure support mechanism. For gravitational condensations after a gravitational free-fall time $\tau_{FF} = (\rho G)^{-1/2}$ the density increases exponentially to form a planet or star with internal stresses to bring the condensation to a halt and allow hydrostatic equilibrium with pressure forces in balance with gravitational forces [5]. For gravitational void formation $\tau_{FF} = (\rho G)^{-1/2}$ is the time required for the rarefaction wave of the growing void to reach its limiting speed at Mach 1 from the Rankine-Hugoniot relations, since rarefaction shocks are impossible from the second law of thermodynamics. This permits the prediction by HGD of the dominant size of CMB temperature anisotropies (Fig. 1a) to be $ct / 3^{1/2}$ or $1.7 \times 10^{21}$ m at time $t = 10^{13}$ s, as observed.

The density and rate-of-strain of the plasma at transition to gas at 300,000 years ($10^{13}$ s) are preserved as fossils of the time of first structure at 30,000 years ($10^{12}$ s), as shown in Fig. 1e. The plasma turbulence is weak at transition, so the Schwarz viscous scale $L_{SV} = (\gamma\nu / \rho G)^{1/2}$ and Schwarz turbulence scale $L_{ST} = \varepsilon^{1/2} / (\rho G)^{3/4}$ are nearly equal, where $\gamma$ is the rate-of-strain, $\nu$ is the kinematic viscosity, $\varepsilon$ is the viscous dissipation rate and $\rho$ is the density. Because the temperature, density, rate-of-strain, composition and thus kinematic viscosity of the primordial gas are all well known it is easy to compute the fragmentation masses to be that of protogalaxies composed almost entirely of $10^{24} - 10^{25}$ kg planets in million-solar-mass $10^{36}$ kg (PGC) clumps [4]. The NBDM diffuses to diffusive Schwarz scales $L_{SD} = (D^2 / \rho G)^{1/4}$ much larger than $L_N$ scale protogalaxies, where $D$ is the NBDM diffusivity and $D \gg \nu$. The rogue-planet prediction of HGD was promptly and independently predicted by the Schild 1996 interpretation of his quasar microlensing observations [14].

The HGD prediction of $\mathrm{Re}_\lambda \sim 100$ weak turbulence in the plasma epoch is supported by statistical studies of cosmic microwave background (CMB) temperature anisotropy fine structure compared to atmospheric, laboratory and numerical simulation turbulence values [19-21]. Plasma turbulence imposes a preferred direction to the massive plasma objects formed by gravitational instability in the plasma at length scales that reflect fossil temperature turbulence of the big bang [6]. Baroclinic torques produce vorticity at rate $\partial\vec{\omega} / \partial t = \nabla\rho \times \nabla p / \rho^2$ at the boundary of protosupercluster voids as shown in Fig. 1e [11]. Because $\nabla\rho$ is roughly constant over $L_{SF\,\mathrm{bigbang}}$ fossil turbulence scales of the big bang turbulence at strong force freeze-out stretched by inflation, HGD explains observations of quasar polarization matching the direction of the Axis of Evil to Gpc scales approaching the present horizon scale $L_H = ct$ [22]. This direction on the cosmic sphere is right ascension RA = $202^{\circ}$, declination $\delta = 25^{o}$, which matches the 2-4-8-16 directions of CMB spherical harmonics [23] and galaxy spins to supercluster 30 Mpc scales [ 24], and is quite unexpected and inexplicable from ΛCDM theory. All temperature fluctuations of big bang turbulence are fossilized by inflation, which stretches space and the fluctuations beyond the scale of causal connection $ct$ at speeds $\approx 10^{25} c$ [6].

## 3. Observations

Figure 2 (top) shows the Tadpole (VV29, UGC 10214) galaxy merger system imaged by the HST/ACS camera, compared to a Keck Telescope spectroscopic study (bottom) by Tran et al. 2003 [23]. The galaxy dark matter clearly consists of PGCs since the spectroscopy proves the YGCs were formed in place in the galaxy halo and not ejected as a collisionless tidal tail [20]. Quasar microlensing [21] suggests the dark matter PGCs must be composed of frozen planets in metastable equilibrium, as predicted by HGD [11].

Figure 2 (bottom) shows a linear trail of YGCs pointing precisely to the frictional spiral merger of the small blue galaxy VV29c that is embedded in the accretion disk of VV29a, Fig. 2 (top). With tidal agitation from the merger, the planets undergo an accretional cascade to larger and larger planets, and finally form stars within the PGCs that are stretched away by tidal forces to become field stars in the observed star-wakes and dust-wakes. All galaxies originate as protogalaxies at 0.03 Mpc scales $L_N$ reflecting viscous-gravitational fragmentation of weakly turbulent plasma just before its transition to gas. A core of 13.7 Gyr old stars at scale $L_N$ persists in most if not all galaxies bound by PGC-viscosity of its remaining PGC dark matter. Most of the PGC mass diffuses out of the protogalaxy core to form the galaxy dark matter halo, observed to extend to a diameter of 0.3 Mpc in Tadpole, Fig. 1 (top). A more detailed discussion is found elsewhere [20].

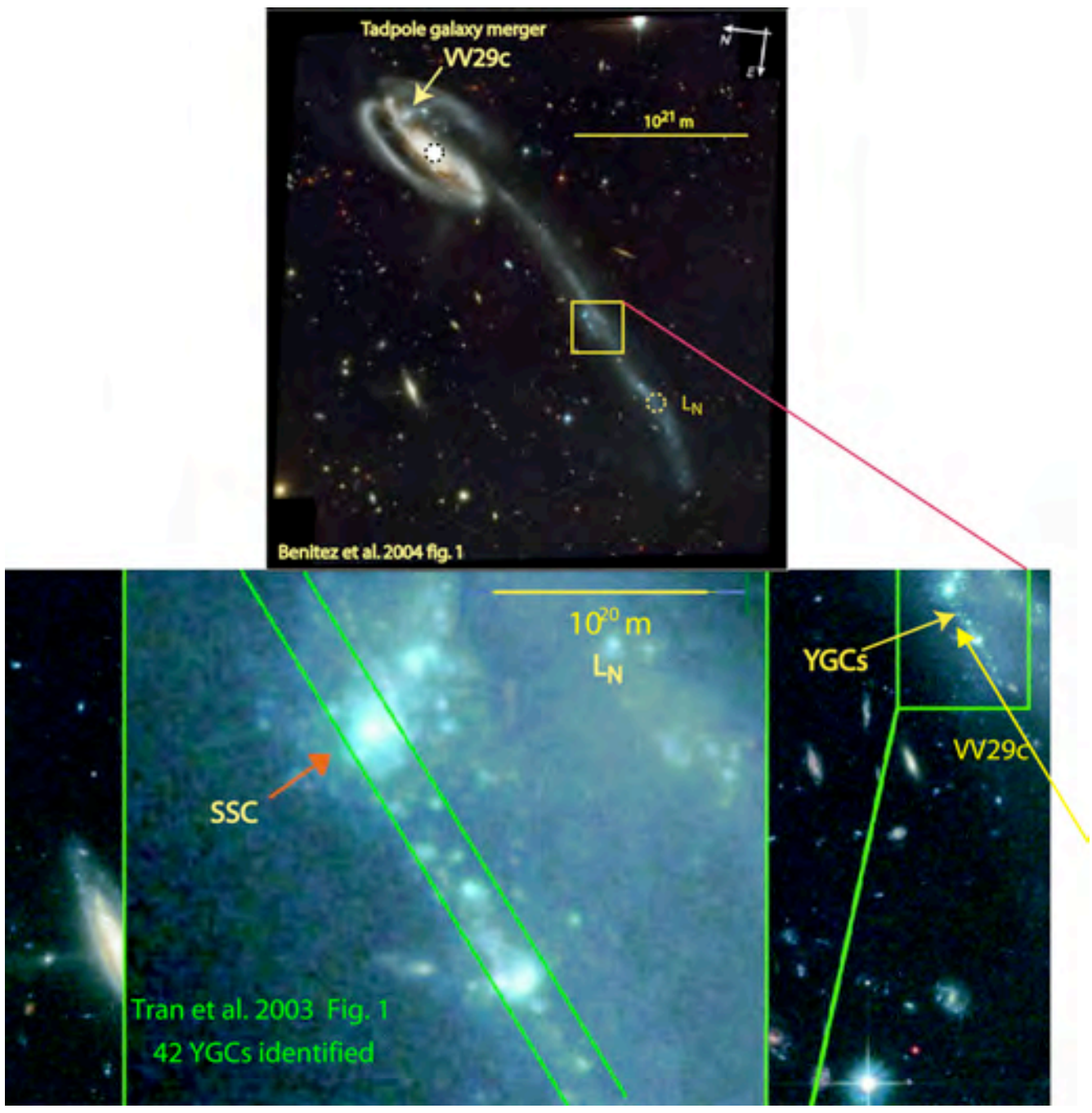

Fig. 2. Tadpole galaxy merger system illustrating the size of the baryonic dark matter system surrounding the central galaxy VV29a and the frictional spiral merger of galaxy VV29c leaving a trail of young-globular-star-clusters (YGCs)

Figure 3 shows a planetary nebula in the Large Magellanic Cloud (LMC) claimed from a recent brightness episode to be on the verge of a SNe Ia event [32], where the central white dwarf and companion (or possibly just a white dwarf and a JPP accretion disk) have ejected several solar masses of matter in the bright clumps observed. The HGD interpretation is very different [17]. From HGD the bright clumps are formed in place from clumped baryonic-dark-matter frozen planets termed JPPs (Jovian PFP Planets of all sizes form by gassy binary accretional mergers of PFPs and their growing daughters), where some of the multi-Jupiter-mass clumps (globulettes [33]) are accreted by the star, and none are ejected. As the JPPs are accreted the star shrinks, its mass and density increase, and its spin rate increases producing a powerful plasma beam that evaporates the frozen gas planets it encounters.

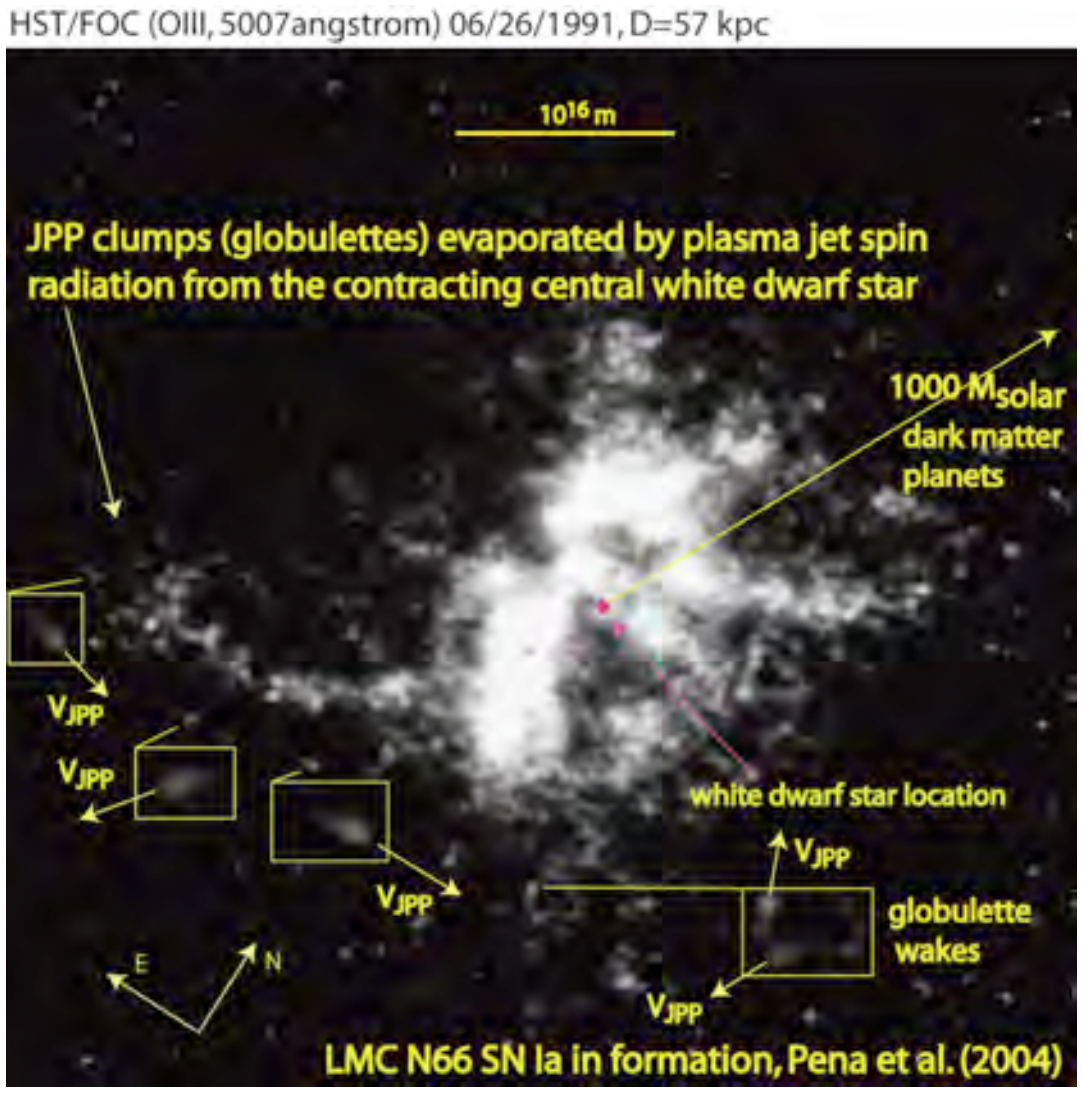

Fig. 3. Planetary nebula LMC N66 suggested by Pena et al. 2004 as a Supernova Ia in formation [25]. From HGD the HST/FOC image reveals globulette clumps [26] of dark matter planets evaporated by the plasma jet of the rapidly spinning, contracting, white dwarf star at the center, burdened by the rain of accreting, evaporating, planets.

Bright wakes can be seen for JPP-PFP globulettes (multi-Jupiter mass planet clumps) in the Fig. 3 magnified images (boxes) showing the objects are moving in random directions with speeds $V_{JPP}$. Presumably the globulette speed determines the rate of accretion by the central star and the rate of its radial mixing of thermonuclear products [10]. Slow speeds will reduce the size of the JPPs and their clumps, and increase the probability that the central star will die quietly as a helium white dwarf with mass < 1.4 solar. Moderate JPP accretion rates may fail to mix away the carbon core giving a supernova Ia at the Chandresekhar critical mass 1.44 solar. Stronger mixing and accretion may permit an iron core and a supernova II event. Even stronger accretion rates lead to superstars or black holes. Within the PNe size there should be more than a thousand solar masses of dark matter planets, as shown by the arrow toward upper right.

Figure 4 shows the Helix planetary nebula. Helix is much closer (209 pc) and presumably less strongly agitated by tidal forces from other objects than the LMC N66 PNe of Fig. 3.

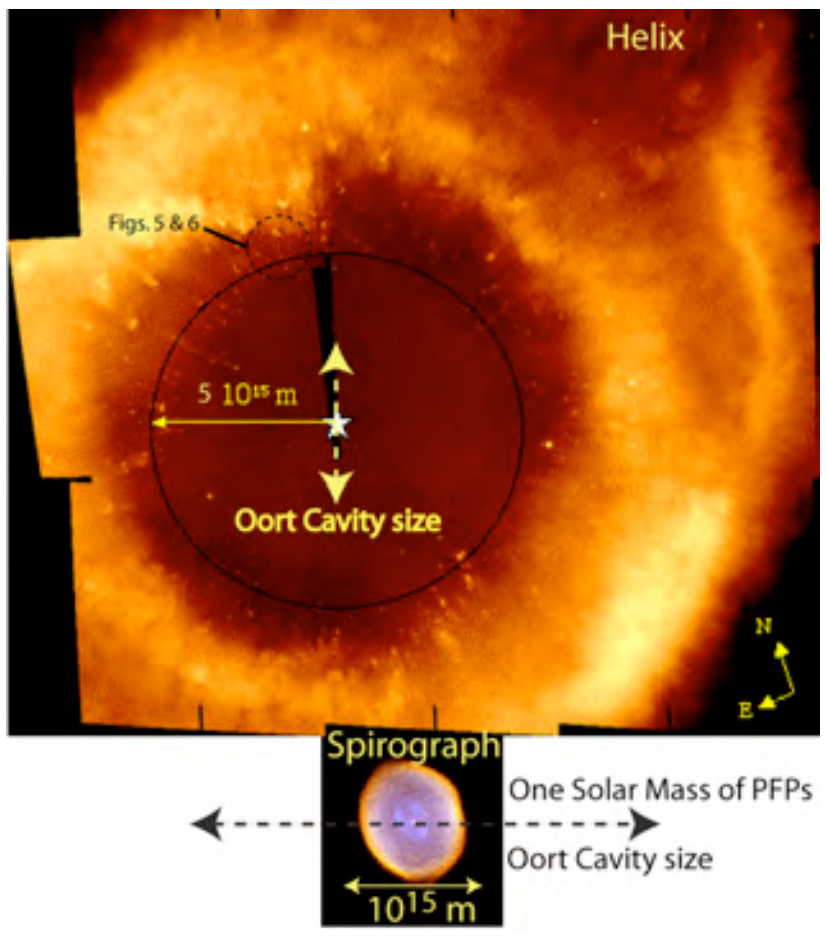

Fig. 4. Helix planetary nebula (top) showing numerous dark matter planets evaporated by the central white dwarf at the outer boundary of the Oort cavity left by the accretion of PFPs to form the star within the PGC. Spirograph PNe (bottom) is young and still growing within its Oort cavity, where apparently no dark matter planets remain unevaporated. Detailed images of Helix dark matter planets are shown in Fig. 5 and Fig. 6 at the indicated location north of the central star.

Thousands of evaporating gas planets can be seen in Fig. 4 (and Fig. 5 and Fig. 6 close-up) HST images. The planets have spacing consistent with the fossil density from the time of first structure at 30,000 years after the big bang; that is, $\rho_0 \sim 10^{-17}$ kg m$^{-3}$.

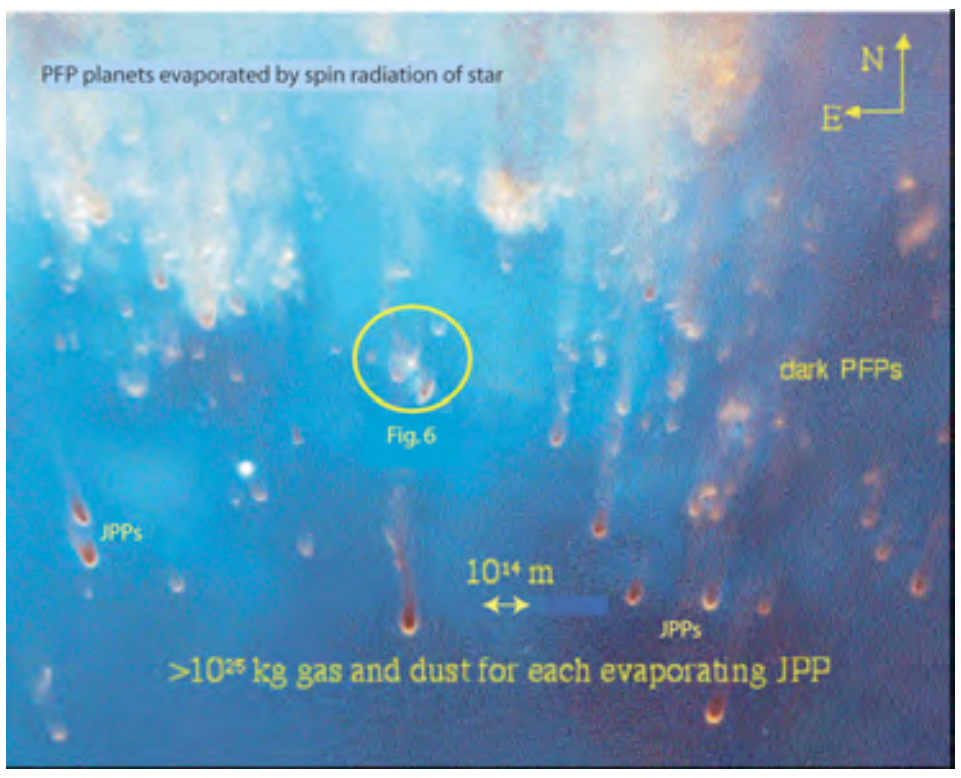

Fig. 5. Closeup image of the region north of the central star shown by the dashed circle in Fig. 4. The central white dwarf spins rapidly because it has shrunk to a density of order $10^{10}$ kg m$^{-3}$ from the mass of accreted planets. A bipolar plasma beam irradiates and evaporates dark matter planets at the edge of the Oort cavity, creating the planetary nebula.

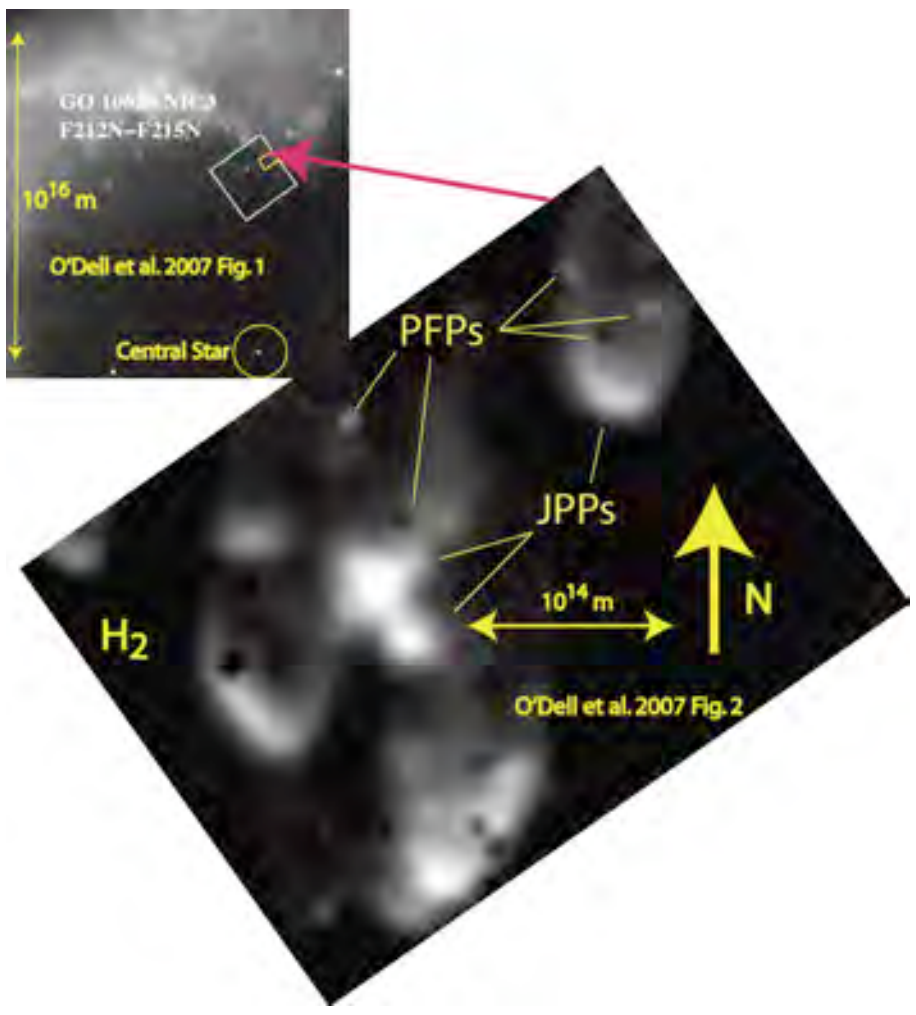

Fig. 6. Detail of location north of central white dwarf star in Helix PNe [35], showing evaporating PFP planets as well as large JPP planets and their atmospheres that can dim a SNe Ia event if it is along the line of sight, Fig. 8.

The mass of white-dwarf-precursor stars are vastly overestimated by ignorance of the existence of dark matter planets. Without PFPs, standard PNe models assume the mass of the nebula originates in the original star and is somehow ejected as clumps and superwinds. Figure 6 shows close-up images of the JPP and partially evaporated PFP planets, illuminated by the spin-radiation of the central star.

Figure 7 shows estimates of the final mass of stars $M_{Final} \leq 1.3 M_{sun} \leq M_{Pulsar} \leq M_{SNeIa}$ compared to the initial mass $M_{PNE}$ where the PNe mass is estimated from the nebulae brightness assuming all of the nebula mass originates in the star and none from dark matter planets.

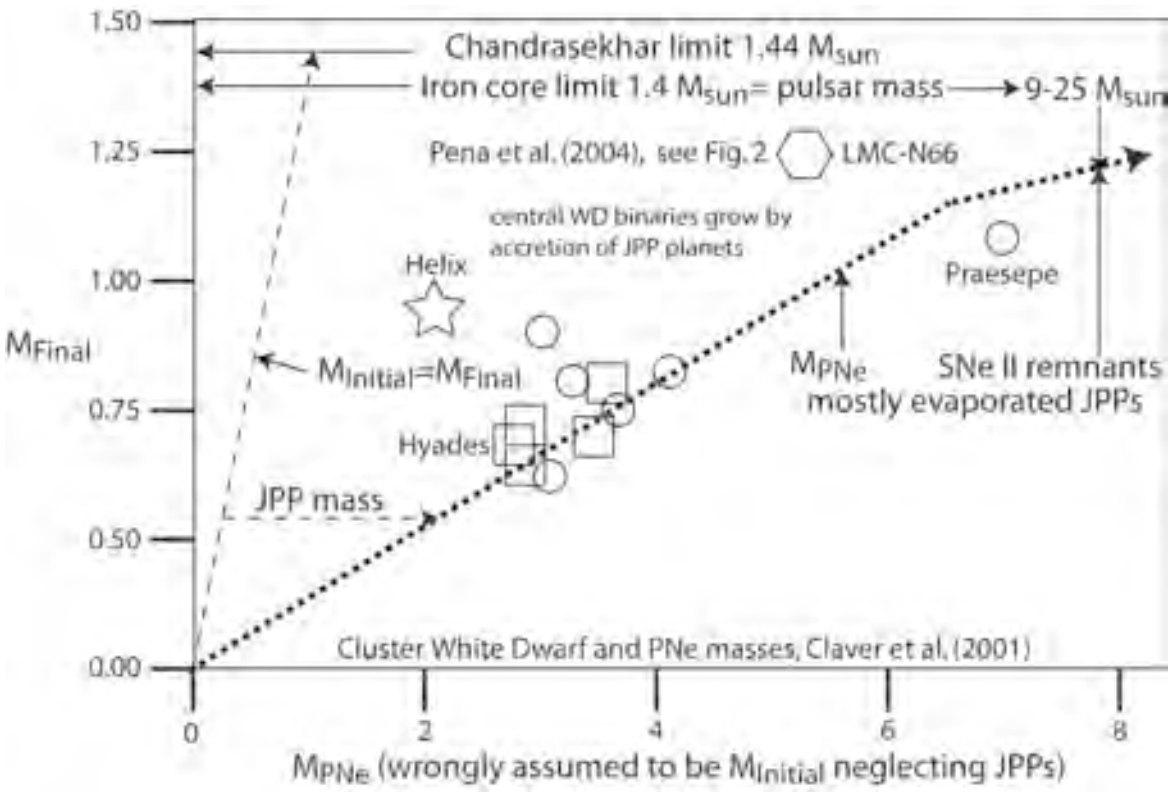

Fig. 7. Star masses compared to assumed PNe and SNe initial star masses neglecting dark matter planet effects such as JPP atmosphere brightness and the mass of evaporated dust and gas [17].

We see from Fig. 7 that final star masses are much smaller than initial masses by standard models of PNe and SNe. Pulsar masses $M_{Pulsar} = 1.4 M_{sun}$ are less than the Chandrasekhar white dwarf limit $M_{Chandrasekhar} = 1.44 M_{sun}$ .

Figure 8 shows that an alternative to dark energy is dark matter planets. Fig. 8 (top) is dimness of SNe Ia events vesus redshift z for uniform-grey-dust, dark-energy (nonlinear-grey-dust) and no-dark-energy models. The uniform-grey-dust model fails at large z. Dark matter planets provide a nonlinear-grey-dust effect if SNe Ia events take place in PNe surroundings such as the Helix, Fig. 8 (bottom) where spin-radiation of a white dwarf creates JPP atmospheres that may dim or not dim depending on the line of sight.

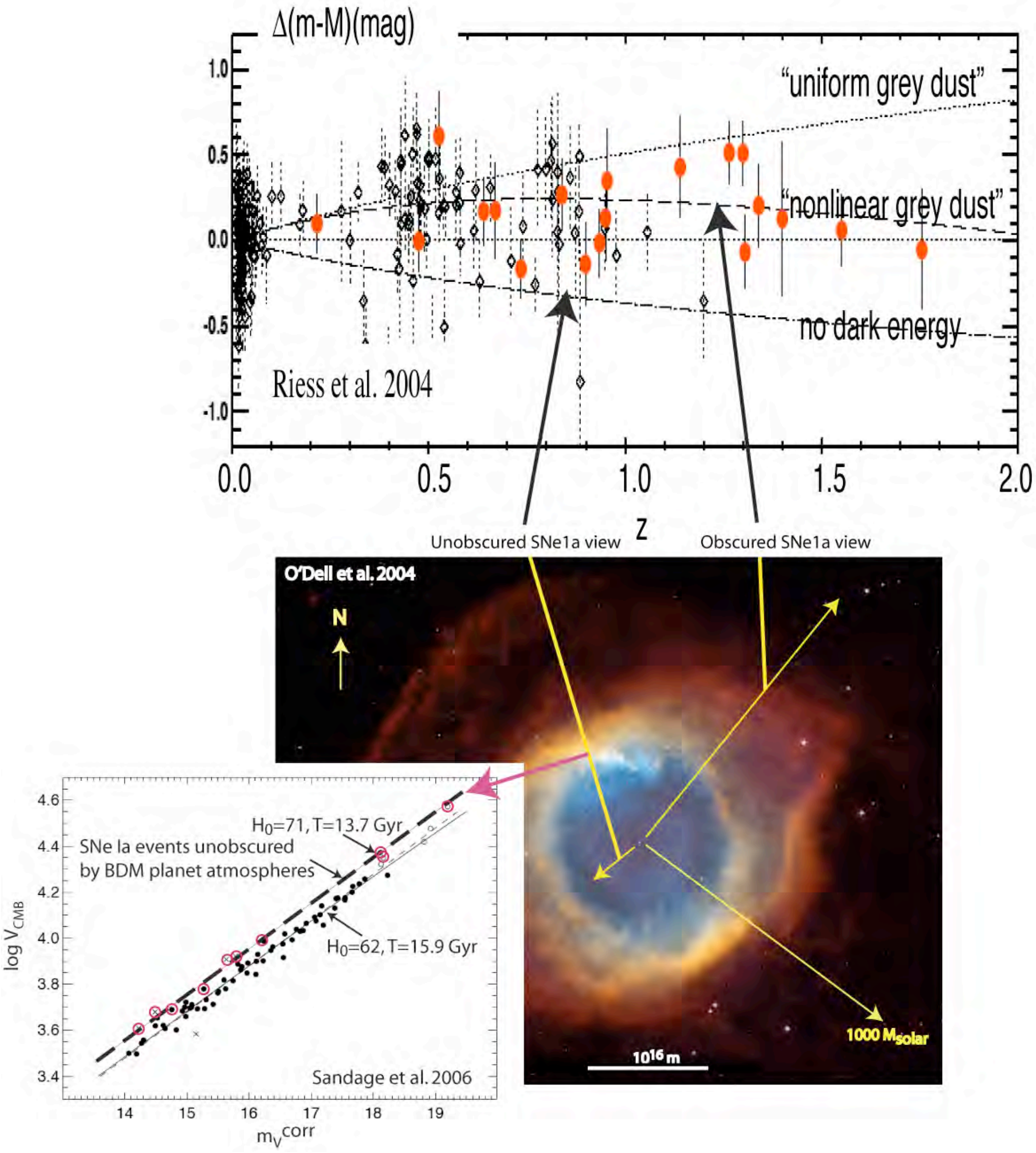

Fig. 8. Helix planetary nebula showing the effect of JPP planetary atmospheres along the line of sight to SNe Ia events is to produce a systematic dimming error that can masquerade both as dark energy (top) or as increased Hubble constants and ages of the universe (left insert). Open circles (red) are bright SNe Ia events of the Sandage 2006 data set taken to be unobscured by evaporated planetary atmospheres, and supporting the CMB universe age T = 13.7 Gyr.

The brightest SNe Ia events agree with the no-dark-energy curve of Fig. 8 (top) and can be interpreted as lines of sight that do not intersect dense dark matter planet atmospheres. A similar interpretation is given to the Sandage 2006 [27] SNe Ia global Cepheid Hubble-Constant $H_0$ = 62.3 km $s^{-1}$ $Mpc^{-1}$ estimates that disagree with the WMAP $H_0$ = 71 km $s^{-1}$ $Mpc^{-1}$ value and estimate of the age of the universe T to be 15.9 Gyr rather than the CMB value of 13.7 Gyr. Taking the least dim SNe Ia values measured to be correct removes the systematic error of dark matter planet atmosphere dimming (non-linear grey dust), so discrepancies in T and $H_0$ are removed.

Figure 9 shows luminosity scales for gamma-ray-burst GRB events extrapolated with the SNe Ia events of Fig. 8 top. GRB power can exceed that of $10^{23}$ stars or a trillion galaxies, permitting detection at redshifts z>5.8. Such events imply the formation of dense central objects in protogalaxies, contrary to ΛCDM cosmology where galaxies should not be formed at such redshifts, let alone central objects. At large redshifts luminosity distances can exceed Hubble distances depending on the cosmology assumed, but the GRB luminosity distances exceed maximum possible ratios for any cosmology. Squares and a dashed line indicate maximum $D_L / L_H$ for a flat universe with no $\Lambda$-dark-energy. The data suggest both the SNe Ia events and the GRB events have been dimmed by evaporated planets-in-clumps surrounding the central powerful events. Small values below the curve suggest clear lines of sight and a slightly closed universe, as expected from necessary frictional losses of big bang turbulence [14].

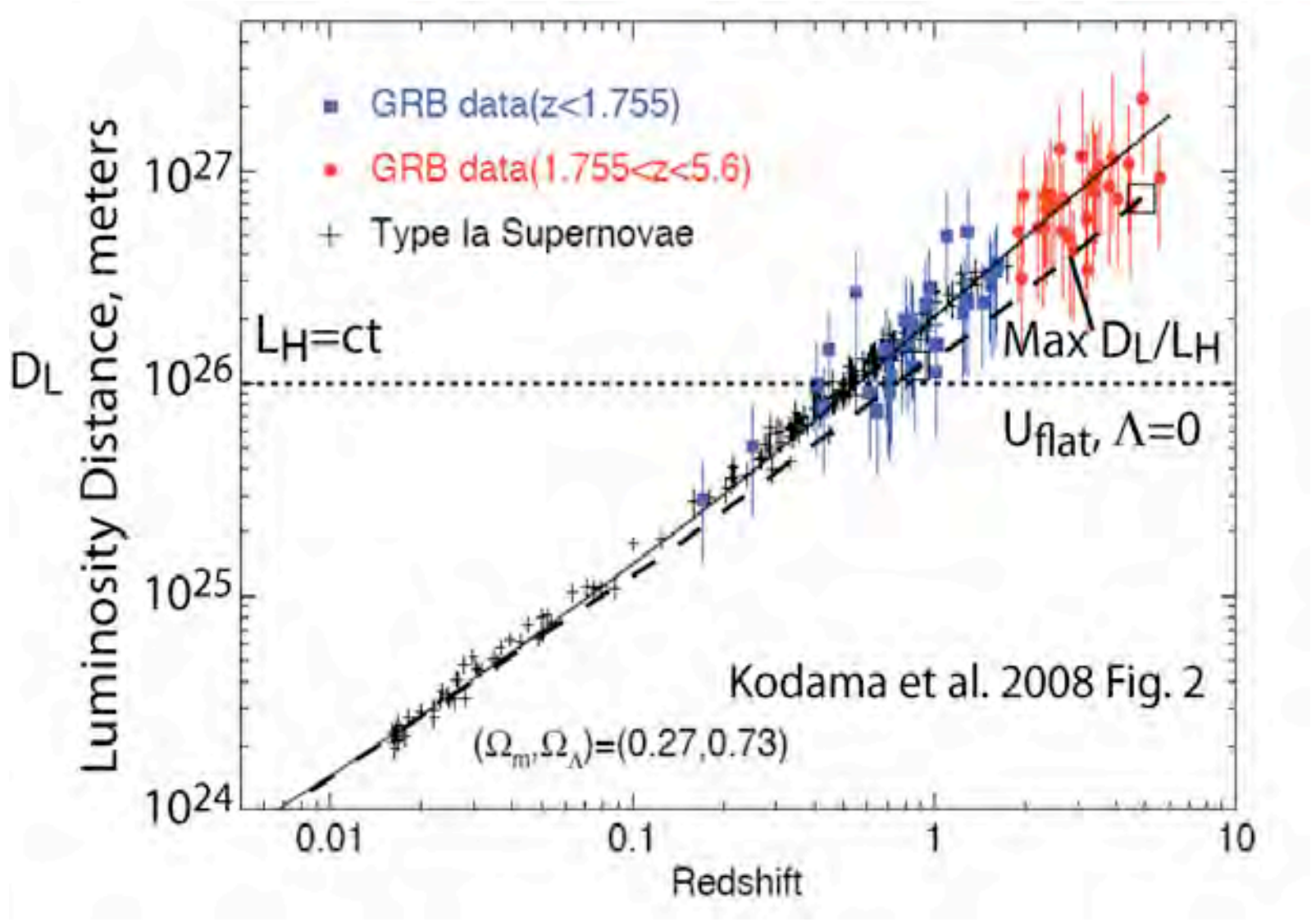

Fig. 9. Gamma ray burst data of Kodama et al. 2008 [36] are calibrated with SNe Ia data of Fig. 8. Luminosity distance scales are larger than physically possible values because both the SNe Ia events and the GRB events are dimmed by intervening partially evaporated galaxy-dark-matter planets-in-clumps. The dashed line with squares indicate the maximum possible $D_L / L_H$ ratio for a flat universe with no dark energy or $\Lambda$.

## 4. Discussion

An accumulation of evidence in a variety of frequency bands from a variety of very high resolution and highly sensitive modern telescopes leaves little doubt that the dark matter of galaxies is primordial planets in proto-globular-star-cluster clumps, as predicted from HGD by Gibson 1996 [11] and inferred from quasar microlensing by Schild 1996 [21]. All stars form from these planets so all star models and planetary nebulae models must be revised to take the effects of planets and their brightness and dimness effects into account. Turbulence produces post-turbulence (fossil turbulence) with structure in patterns that preserve evidence of previous events such as big bang turbulence and plasma epoch turbulence. Post-turbulence perturbations [6] guide the evolution of all subsequent gravitational structures. Numerous fatal flaws in the standard ΛCDMHC cosmology have appeared that can be traced to inappropriate and outdated fluid mechanical assumptions [22] that can be corrected by HGD [11-21]. A critically important advance in the understanding of fluid mechanics is the correct definition of turbulence and the realization that all turbulence cascades from small scales to large [4-10]. Failure to recognize this advance has hampered the field of oceanography, which is therefore blind to the importance of fossil turbulence, fossil turbulence waves, zombie turbulence and zombie turbulence waves as the dominant physical mechanism of vertical transport of hydrophysical fields and for the preservation of information about previous turbulence [37-39].

## 5. Conclusions

Dark matter planet dimming errors account for the SNe Ia overestimate (T=15.9 Gyr) of the age of the universe [27] and dark energy dimming of SNe Ia events, as described by Fig. 8. Dark energy is an unecessary and incorrect hypothesis from HGD. Thus we need not modify any physical laws nor predict the end of cosmology because evidence of cosmological beginnings are being swept out of sight by an accelerating vacuum-antigravity-powered expansion of the universe [40,41].

From HGD and the second law of thermodynamics, the frictional non-adiabatic big bang turbulence beginning of the universe [11] implies the universe is closed, not open. Because the big bang was extremely hot, the big bang turbulence produced little entropy so the departures from a flat universe should be small, as observed. Fossil strong-force freeze-out scale density turbulence explains the Gpc scales of quasar polarization vectors in alignment with the CMB axis of evil [29, 31, 18], as shown by the dashed circles in Fig. 1. Negative big bang turbulence stresses and negative gluon viscous stresses are candidates to drive the exponential inflation of space during and at the end of the big bang. An inflationary event is indicated by observations showing similar galaxy patterns exist in regions outside each others causal connection scales.

Star formation models and planetary nebulae formation models must be corrected to account for the effects of dark matter planets [24]. Initial star masses have been vastly overestimated from the brightness of evaporating planets formed around dying central stars by spin radiation, as shown by Fig. 7. Red giant and asymptotic giant branch (AGB) excursions of brightness on the Hertzsprung-Russell color-magnitude diagram for star evolution must be reexamined to account for the fact that all stars are formed and grow by the accretion of planets that have first order effects on both the color and magnitude of stars as they evolve. Planetary nebulae are apparently not dust clouds ejected by AGB superwind events as usually assumed but are partially evaporated and brightly illuminated dark matter planets, as shown by Figs. 3-6, and Fig. 8.

## References

1. Riess, A. G., Filippenko, A.V. et al. 1998. Observational evidence from supernovae for an accelerating universe and a cosmological constant AJ, 116, 1009.
2. Perlmuter, S., Aldering, G. et al. 1999. Measurements of $\Omega$ and $\Lambda$ from 42 high-redshift supernovae ApJ, 517, 565.
3. Chernin, A.D., Karachentsev, I.D. et al. 2007. Detection of dark energy near the Local Group with the Hubble Space Telescope, arXiv:astro-ph/0706.4068v1.
4. Gibson, C.H. (1991). Kolmogorov similarity hypotheses for scalar fields: sampling intermittent turbulent mixing in the ocean and galaxy, Proc. Roy. Soc. Lond. A, 434, 149-164.
5. Gibson, C. H. (2006). Turbulence, update of article in Encyclopedia of Physics, R. G. Lerner and G. L. Trigg, Eds., Addison-Wesley Publishing Co., Inc., pp.1310-1314.
6. Gibson, C. H. (1981). Buoyancy effects in turbulent mixing: Sampling turbulence in the stratified ocean, AIAA J., 19, 1394.
7. Gibson, C. H. (1968a). Fine structure of scalar fields mixed by turbulence: I. Zero-gradient points and minimal gradient surfaces, Phys. Fluids, 11: 11, 2305-2315.
8. Gibson, C. H. (1968b). Fine structure of scalar fields mixed by turbulence: II. Spectral theory, Phys. Fluids, 11: 11, 2316-2327.
9. Gibson, C. H. (1986). Internal waves, fossil turbulence, and composite ocean microstructure spectra," J. Fluid Mech. 168, 89-117.
10. Gibson, C. H. (1999). Fossil turbulence revisited, J. of Mar. Syst., 21(1-4), 147-167, astro-ph/9904237
11. Gibson, C.H. (1996). Turbulence in the ocean, atmosphere, galaxy and universe, Appl. Mech. Rev., 49, no. 5, 299–315.
12. Gibson, C.H. (2000). Turbulent mixing, diffusion and gravity in the formation of cosmological structures: The fluid mechanics of dark matter, J. Fluids Eng., 122, 830–835.
13. Gibson, C.H. (2004). The first turbulence and the first fossil turbulence, Flow, Turbulence and Combustion, 72, 161–179.
14. Gibson, C.H. (2005). The first turbulent combustion, Combust. Sci. and Tech., 177: 1049–1071, arXiv:astro-ph/0501416.
15. Gibson, C.H. (2006). The fluid mechanics of gravitational structure formation, astro-ph/0610628.
16. Gibson, C.H. (2008). Cold dark matter cosmology conflicts with fluid mechanics and observations, J. Applied Fluid Mech., Vol. 1, No. 2, pp 1-8, 2008, arXiv:astro-ph/0606073.
17. Gibson, C.H. & Schild, R.E. (2007). Interpretation of the Helix Planetary Nebula using Hydro-Gravitational-Dynamics: Planets and Dark Energy, arXiv:astro-ph/0701474.
18. Schild, R.E & Gibson, C.H. (2008). Lessons from the Axis of Evil, axXiv[astro-ph]:0802.3229v2.
19. Gibson, C.H. & Schild, R.E. (2007). Interpretation of the Stephan Quintet Galaxy Cluster using Hydro-Gravitational-Dynamics: Viscosity and Fragmentation, arXiv[astro-ph]:0710.5449.
20. Gibson, C.H. & Schild, R.E. (2002). Interpretation of the Tadpole VV29 Merging Galaxy System using Hydro-Gravitational Theory, arXiv:astro-ph/0210583.
21. Schild, R. 1996. Microlensing variability of the gravitationally lensed quasar Q0957+561 A,B, ApJ, 464, 125.
22. Jeans, J. H. 1902. The stability of spherical nebula, Phil. Trans., 199A, 0-49.
23. Tran, H. D., Sirianni, M., & 32 others 2003. Advanced Camera for Surveys Observations of Young Star Clusters in the Interacting Galaxy UGC 10214, ApJ, 585, 750.
24. Toomre, A., & Toomre, J. 1972. Galactic Bridges and Tails, ApJ, 178, 623.

25. Gibson, C.H. & Schild, R.E. (2002). Interpretation of the Tadpole VV29 Merging Galaxy System using Hydro-Gravitational Theory, arXiv:astro-ph/0210583.
26. Bershadskii, A. 2006. Isotherms clustering in cosmic microwave background, Physics Letters A, 360, 210-216.
27. Bershadskii, A., and Sreenivasan, K.R. 2002. Multiscaling of cosmic microwave background radiation, Phys. Lett. A, 299, 149-152.
28. Bershadskii, A., and Sreenivasan, K.R. 2003. Extended self-similarity of the small-scale cosmic microwave background anisotropy Phys. Lett. A, 319, 21-23.
29. Hutsemekers, D. et al. 2005. Mapping extreme-scale alignments of quasar polarization vectors, A&A 441, 915–930.
30. Land, K. & Magueijo, J. 2005. Phys. Rev. Lett. 95, 071301.
31. Longo, M. J. 2007. Evidence for a Preferred Handedness of Spiral Galaxies arXv:0707.3793.
32. Pena, M. et al. 2004. A high resolution spectroscopic study of the extraordinary planetary nebula LMC-N66, A&A, 419, 583-592.
33. Gahm, G. et al. 2007. Globulettes as seeds of brown dwarfs and free-floating planetary-mass objects, AJ, 133, 1795-1809.
34. Sandage, A. et al. 2006. The Hubble constant: a summary of the HST program for the luminosity calibration of Type Ia supernovae by means of Cepheids, ApJ, 653, 843.
35. O'Dell, C.R., Henney, W. J. & Ferland, G. J. 2007. Determination of the physical conditions of the knots in the Helix nebula from optical and infrared observations, AJ, 133, 2343-2356, astro-ph/070163.
36. Kodama, Y., Yonetoku, D. et al. 2008. Gamma-ray bursts in $1.8<z<5.6$ suggest that the time variation of the dark energy is small, for MNRAS, arXiv0802.3428.
37. Keeler, R. N., V. G. Bondur, and C. H. Gibson 2005. Optical Satellite Imagery Detection of Internal Wave Effects from a Submerged Turbulent Outfall in the Stratified Ocean, Geophysical Research Letters, Vol. 32, L12610.
38. Gibson, C. H., Bondur, V. G., Keeler, R. N. and Leung, P. T. (2006). Remote sensing of submerged oceanic turbulence and fossil turbulence, International Journal of Dynamics of Fluids, Vol. 2, No. 2, 171-212.
39. Gibson, C. H., Bondur, V. G., Keeler, R. N. and Leung, P. T. (2008). Energetics of the beamed zombie turbulence maser action mechanism for remote detection of submerged oceanic turbulence, Journal of Applied Fluid Mechanics, Vol. 1, No. 1, 11-42.
40. Krauss, L. M. and Scherrer, R. J. 2008. The end of cosmology, Scientific American, 298 (3), 46-53.
41. Krauss, L. M. and Scherrer, R. J. 2007. The return of a static universe and the end of cosmology, J. Gen. Rel. and Grav., 39, 10, 1545-1550, arXv:0704.0221.